\documentclass[12pt,preprint]{revtex4-1}

\usepackage[]{graphicx}
\usepackage{amsmath,amssymb}
\usepackage{times}

\begin{document}

\title{Spectrum and mass composition of cosmic rays in the energy range $10^{15}-10^{18}$~eV derived from the Yakutsk array data}

\author{S.\,P.~Knurenko}\email{s.p.knurenko@ikfia.ysn.ru}
\author{A.~Sabourov}

\affiliation{Yu.\,G.~Shafer Institute of cosmophysical research and aeronomy SB RAS 677980 Lenin Ave. 31, Yakutsk, Russia}

\newcommand{\etal}{\MakeLowercase{\textit{et~al.}}}
\newcommand{\xmax}{x_{\text{max}}}
\newcommand{\seff}{s_{\text{eff.}}}
\newcommand{\MeanLnA}{\left<\ln{\text{A}}\right>}

\begin{abstract}
  A spectrum of cosmic rays within energy range $10^{15} - 3 \times 10^{17}$~eV was derived from the data of the small Cherenkov setup, which is a part of the Yakutsk complex EAS array. In this, work a new series of observation is covered. These observations lasted from 2000 till 2010 and resulted in increased number of registered events within interval $10^{16} - 10^{18}$~eV, which in turn made it possible to reproduce cosmic ray spectrum in this energy domain with better precision. A sign of a thin structure is observed in the shape of the spectrum. It could be related to the escape of heavy nuclei from our Galaxy. Cosmic ray mass composition was obtained for the energy region $10^{16} - 10^{18}$~eV. A joint analysis of spectrum and mass composition of cosmic rays was performed. Obtained results are considered in the context of theoretical computations that were performed with the use of hypothesis of galactic and meta-galactic origin of cosmic rays.
\end{abstract}

\maketitle

\section{Introduction}

Energy spectrum of cosmic rays (CR) in energy range $3 \times (10^{15} - 10^{18})$~eV could not be studied in detail with compact arrays due to their small acceptance at energy above $10^{17}$~eV. At the same time this area of the spectrum is of a great interest, since local irregularities are manifested there: production of kinks (thin structure at $3 \times 10^{15} - 10^{17}$~eV) arising from non-uniform distribution of heavier CR components in our Galaxy. On the other hand, this effect is smoothed by addition of a new component (of meta-galactic or other origin) to the cosmic ray flux near Earth. As a result, presence/absence of significant irregularities in spectra measured by various compact arrays allows one to speculate on the CR origin and propagation in our Galaxy~\cite{Knurenko2011, Kalmykov2008}.

The Yakutsk array in this sense appears as a unique scientific tool. It is related to medium-sized arrays, capable of effective measuring of cosmic rays flux in a wide energy range ($10^{15} - 10^{19}$~eV). Other important traits of the array are its model-independent technique of energy estimation of extensive air showers (EAS) and the ability to track longitudinal EAS development by detecting the Cherenkov light emission. Factors mentioned above enable adopting the unique method, combining the studies of CR spectrum and mass composition aimed at exploration of astrophysical aspect of cosmic rays~\cite{Knurenko2006, Knurenko2007}.

\section{Methodical issues}

\begin{figure}
  \centering
  \includegraphics[width=0.75\textwidth]{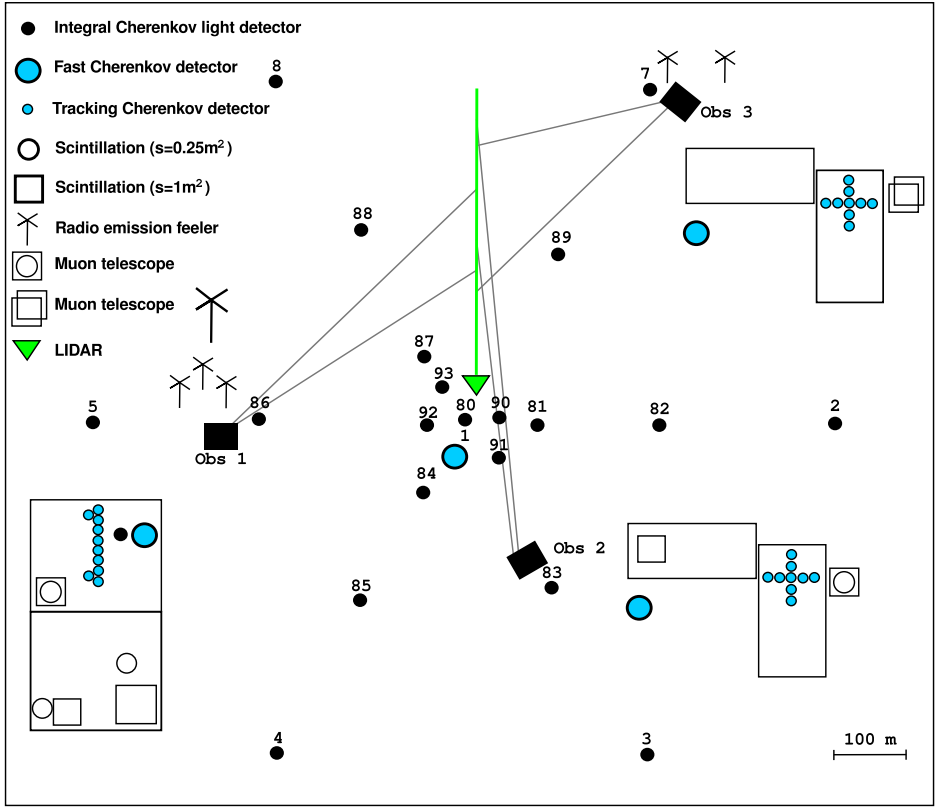}
  \caption{
    Location of observational points of the small Cherenkov setup.
  }
  \label{fig1}
\end{figure}

For more than 15 years the small Cherenkov setup has been operating as a part of the Yakutsk array. It measures Cherenkov light emission from EAS of lower energies (see Fig.\ref{fig1}) using standard detectors which are designed to operate in winter conditions. The area of modern prototype was significantly increased in comparison with the original setup, its border forms a circle of $500$~m radius. The number of optical detectors was also increased (see Fig.\ref{fig1}). Table~\ref{tab1} presents the information on operation of the setup (on annual basis) combined with mean spectral atmosphere transparency at wavelength $430$~nm.

All the information on each shower is stored in the database, which is controlled by a software complex. This program includes units for gathering, sorting and storing of the experimental data. It also includes mathematical units for data processing and statistical analysis. The results of the analysis are presented below.

\begin{table}[ht]
  \caption{
    Characteristics of different observational periods
  }
  \label{tab1}
  \centering
  \begin{tabular}{lllllllll}
    \hline
    {\bf Years} & {\bf 95} & {\bf 95--96} & {\bf 96--97} & {\bf 97--98} & {\bf 98--99} & {\bf 99--00} & {\bf 00--01} & {\bf 01--02} \\
    \hline
    $T$, min & 9960 & 19164 & 22563 & 26586 & 28457 & 29683 & 26530 & 26947 \\
    \hline
    $N$, events & 10956 & 18973 & 27076 & 26055 & 38418 & 27161 & 28652 & 25761 \\
    \hline
    $p$, rel.\,units & 0.61 & 0.59 & 0.62 & 0.58 & 0.64 & 0.60 & 0.62 & 0.61 \\
    \hline
    \hline
    {\bf Years} & {\bf 02--03} & {\bf 03--04} & {\bf 04--05} & {\bf 05--06} & {\bf 06--07} & {\bf 07--08} & {\bf 08--09} & {\bf 09--10} \\
    \hline
    $T$, min & 26912 & 33922 & 29613 & 32044 & 32640 & 29725 & 25514 & 35609 \\
    \hline
    $N$, events & 23548 & 31718 & 30649 & 30634 & 34272 & 31211 & 25131 & 44867 \\
    \hline
    $p$, rel.\,units & 0.59 & 0.61 & 0.63 & 0.61 & 0.63 & 0.63 & 0.61 & 0.65 \\
    \hline
  \end{tabular}
\end{table}

\subsection{Selection, processing and choice of the classification parameter for showers}

In order to reconstruct cosmic ray spectrum we used following criteria for shower selection: 1.~atmospheric transmittance $p_{\lambda} \ge 0.65$; 2.~shower axis must lie within $250$~m from the center of array (for showers with primary energy $E_{0} \le 3 \times 10^{16}$~eV) and within $500$~m (for showers with $E_{0} \ge 10^{16}$~eV); 3.~zenith angle of shower arrival direction $\theta \le 50^{\circ}$; 4.~chance probability of shower detection $w_{\text{r}} \ge 0.9$. Introduction of these criteria was mainly defined by the parameters of detectors used in the experiment (mostly by aperture and thresholds of Cherenkov light detectors) and by atmospheric conditions during optical observation. Since shower events were selected by different triggers, during transition from one trigger to another a threshold effect appears. It manifests itself by sudden local increase of intensity in the spectrum. According to our calculations the magnitude of this effect can reach $\sim 30$\,\%. Technically, the softening of this effect has been achieved by extension of the area controlled by the trigger-50 and by its overlapping with other triggers. For example when triggers coincided, having been activated simultaneously by the same shower. In this case, the number of skipped showers was minimized. Another method is introduction of corrections obtained in simulation of measurement.

According to the criteria described above, the data bank of showers was formed by the parameter $Q(150)$, a density of Cherenkov light flux at $r = 150$~m from shower axis. This parameter was derived from readings of Cherenkov detectors located within $80-200$~m from shower axis. The structure of the data bank was defined by the task~--- it was formed strictly by those periods of observation, which confirmed to adopted shower selection criteria mentioned above. We suppose that chosen conditions are sufficient to avoid distortions related to experimental and methodical errors in reconstruction of cosmic rays energy spectrum.

\subsection{Monitoring of the atmosphere}

\begin{figure}
  \centering
  \includegraphics[width=0.85\textwidth]{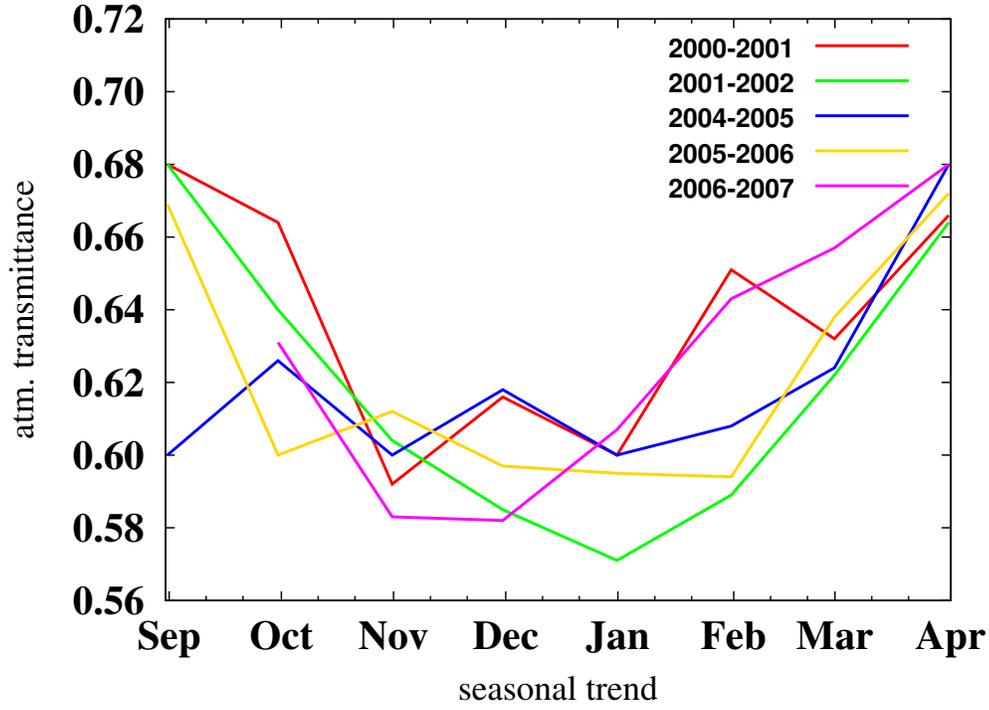}
  \caption{
    Atmospheric transmittance averaged by months for wavelength $430$~nm.
  }
  \label{fig2}
\end{figure}

It is believed that photon losses in clear atmosphere arise from Relay scattering ($5$\,\% from total flux). In real conditions there is  significant loss in received light due to Mi-aerosol of various size. In winter (in the region where array is located the climate is sharp-continental) the atmosphere above the array is non-standard, its parameters change significantly from autumn to winter and vice-versa. According to work~\cite{Dyakonov91} all this factors should be tracked on an operational basis and taken into consideration when analysing different observational periods. On Fig.~\ref{fig2} perennial data are presented on Cherenkov light transmission in atmosphere during different periods of optical observations. These data were utilized during generation of shower samples from which cosmic ray spectrum had been calculated.

\subsection{Estimation of primary energy}

\begin{figure}
  \centering
  \includegraphics[width=0.85\textwidth]{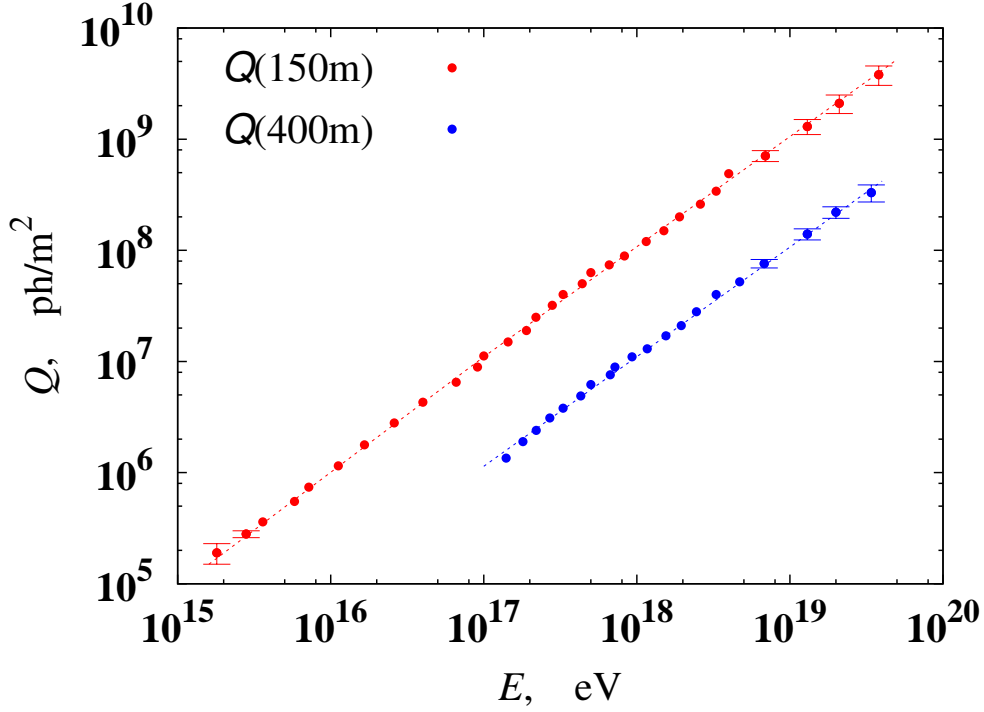}
  \caption{
    Energy dependence of classification parameters.
  }
  \label{fig3}
\end{figure}

Energy was estimated with quasi-calorimetric method, from joint measurements of main shower component. Recently, thanks to significant increase in acceptance for showers with energy $< 500$~PeV, it has become possible for Yakutsk array to measure total number of charged particles on observation level with the precision $(20 \div 30)$\,\% and density of Cherenkov light flux at any given distance from shower axis with the precision $15$\,\%. Such distances for small and large Cherenkov setups are $r = 150$~m and $r = 400$~m accordingly. The densities of Cherenkov light fluxes at these distances were adopted as classification parameters. On fig.~\ref{fig3}, energy dependence of classification parameters is shown. From the data presented on Fig.~\ref{fig3}, a connection between classification parameters and primary energy of a shower $E_{0}$ has been derived:
\begin{equation}
  E_{0} = (9.12 \pm 2.28) \times 10^{16} \cdot \left(\frac{Q(150)}{10^{7}}\right)^{(0.99 \pm 0.02)}
  \label{eq1}
\end{equation}

\begin{equation}
  E_{0} = (8.91 \pm 1.96) \times 10^{17} \cdot \left(\frac{Q(400)}{10^{7}}\right)^{(1.03 \pm 0.02)}
  \label{eq2}
\end{equation}
Expression~(\ref{eq1}) is used in energy range $(5 \div 500)$~PeV and expression~(\ref{eq2})~--- to estimate primary energy of showers above $5 \times 10^{17}$~eV. 

\subsection{Calculation of the $\seff$}

\begin{figure}
  \centering
  \includegraphics[clip, width=0.85\textwidth]{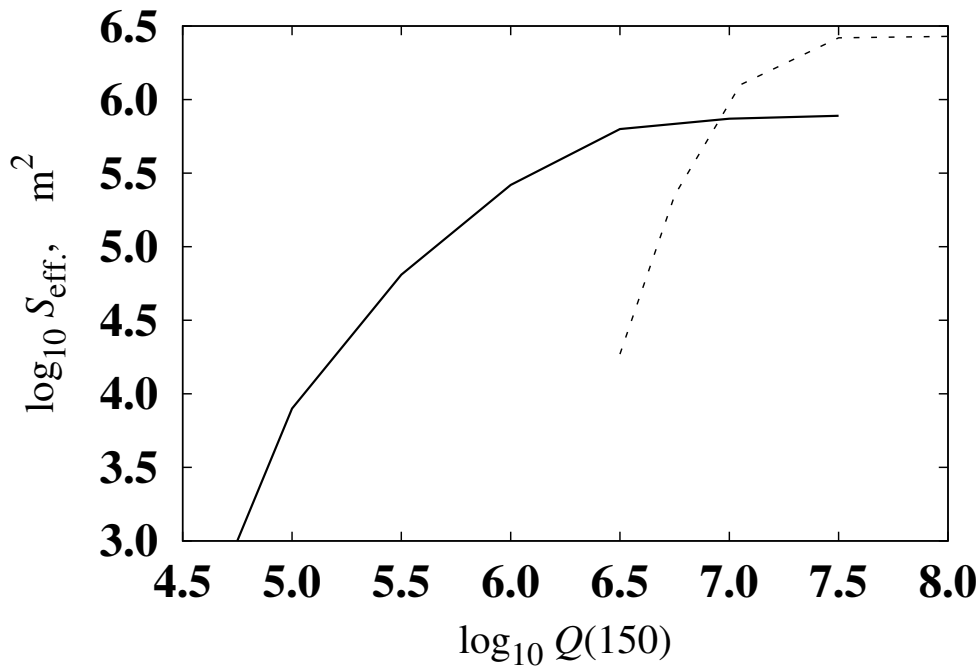}
  \caption{
    Effective area $\seff$ of the small Cherenkov setup for showers from different triggers. Solid line~--- showers with smaller energies (axes within $r = 250$~m), dotted line~--- showers with $E_{0} \ge 10^{17}$~eV (axes within $r = 500$~m).
  }
  \label{fig4}
\end{figure}

The effective area ($\seff$) of the array was taken into consideration for each energy interval according to the method of shower selection. If time of observations and the spatial angle are always defined accurately, then underestimation and overestimation of the $\seff$ usually results in systematic errors in the estimation of the absolute EAS spectrum intensity. This is why the thresholds of Cherenkov detectors ($3 \cdot 10^{5}$~photons/m$^2$) and the effectiveness of shower selection system were considered during the estimation of the $\seff$, i.e. various configurations of triggers of the small setup. For the small Cherenkov array such trigger is C$^{3}$~--- a triple coincidence from Cherenkov detectors. In the latter case a triple excess of the signal amplitude over the night sky background is required. The calculation had been carried out with the use of Monte Carlo method within energy $E_{0}$ intervals with logarithmic step  $\log_{10}{E_{0}} = 15.1$. For each step 1000~artificial showers were thrown with axes distributed within a circle of $250$~m radius for showers of a smaller energies and $500$~m for showers with $E_{0} \ge 10^{17}$~eV (see Fig.~\ref{fig4}). A transition from classification parameters to the energy was made according to empirical formula (\ref{eq1}). When calculating intensity of the spectrum, showers were considered with $\theta \le 50^{\circ}$. The time of the observation is presented in the Table~\ref{tab1}.

\section{Results}

\subsection{Energy spectrum of cosmic rays within energy interval $10^{15} - 10^{18}$~eV}

\begin{figure}
  \centering
  \includegraphics[width=0.85\textwidth, clip]{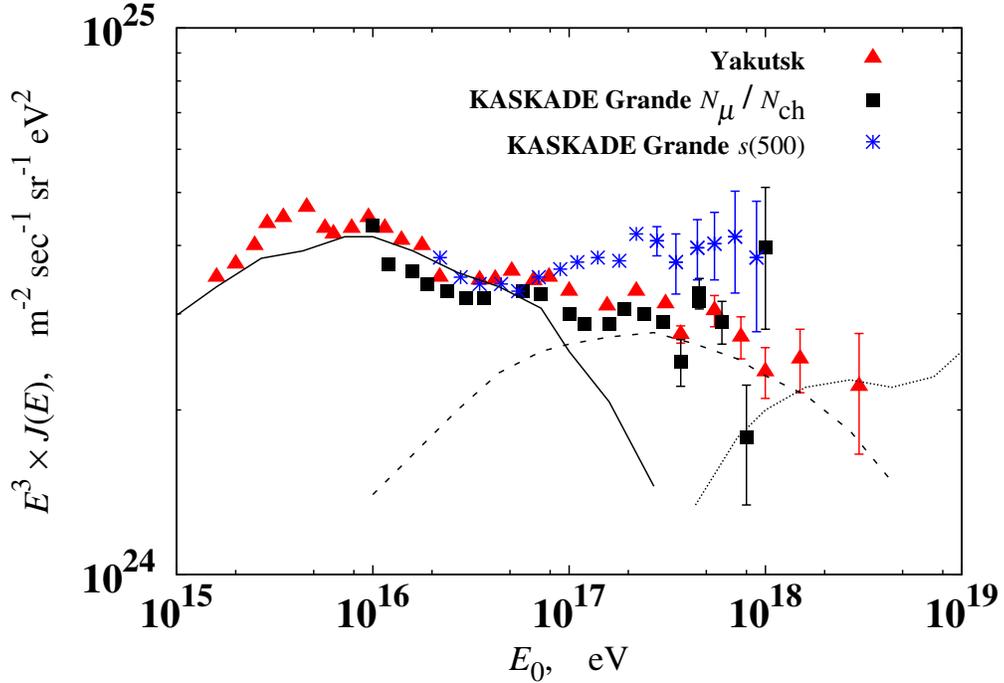}
  \caption{
    Interpretation of the CR spectrum according to scenario~1. The Yakutsk data are denoted with red triangles, black squares and blue stars denote results obtained at KASKADE Grande. Solid line~--- part of the spectrum formed by galactic CR, dotted line~--- extragalactic component, dashed line~--- a component yet to be explained.}
  \label{fig5-1}
\end{figure}

\begin{figure}
  \centering
  \includegraphics[width=0.85\textwidth, clip]{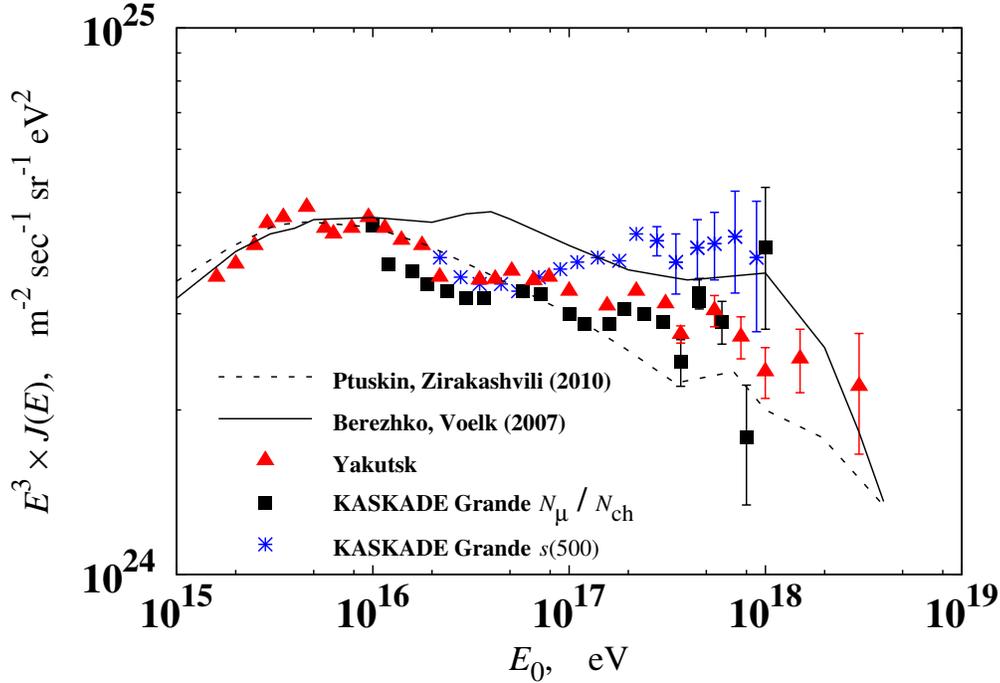}
  \caption{
    Interpretation of the CR spectrum according to scenario~2. Designation of the experimental data is the same as the previous figure. Solid curve~--- theoretical predictions obtained by Berezhko and V\"{o}lk (2009), dotted curve~--- by Ptuskin and Zirakashvili (2010).}
  \label{fig5-2}
\end{figure}

Energy spectrum of cosmic rays with the account of the new data on showers from the selection is presented on Fig.~\ref{fig5-1}-\ref{fig5-2}. Statistical accuracy allows one to speculate of a thin structure within the shape of the spectrum. A comparison with model calculations performed within various hypotheses of CR sources and models of their propagation in the galaxy gives possible interpretation of the experimental data.

Here we consider two possible scenarios of generation of the spectrum. In the scenario~1, the kink at $E = 3 \time 10^{15}$~eV and subsequent increase result from galactic component (up to $10^{17}$~eV). In the region $(\sim 3 \times 10^{17} - 3 \time 10^{18})$~eV, the spectrum is shaped by unknown component; here, one may speculate of CR particles interaction with galactic wind and shock acceleration. In the scenario~2, the galactic component extends to $3 \time 10^{18}$~eV thanks to acceleration by supernova remnants~\cite{Berezhko2007, Ptuskin2010}. This model shows good agreement with experimental data and helps to interpret observed spectrum in the range $10^{15} - 3 \times 10^{18}$~eV.

As it is seen from figures, within the energy range $(5-8) \times 10^{16}$~eV there is a small peak generated by iron nuclei. At lesser energies, where a peak from the CNO group is expected according to the model, a slight increase in the intensity is observed. According to our data and the data from KASCADE-Grande, this exceeding is not significant and could be related to the presence of CR of another origin in the total flux~\cite{Kalmykov2008}.

\subsection{Mass composition}

\begin{figure}
  \centering
  \includegraphics[width=0.85\textwidth, clip]{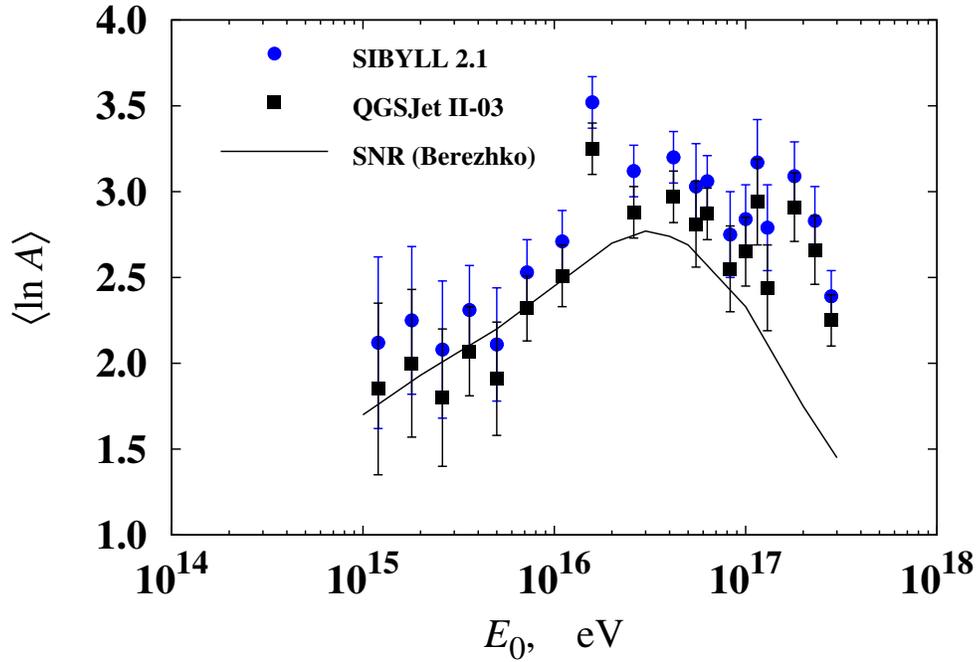}
  \caption{
    The dependence of the $\MeanLnA$ from EAS energy. Predictions of the QGSJet~II-03 model are presented with circles, squares represent mass composition estimation according to the SIBYLL2.1 model. Curve~--- expected composition from supernova remnants (Berezhko and V\"{o}lk, 2009).
  }
  \label{fig6}
\end{figure}

The mean natural logarithm of the CR atomic number $\MeanLnA$ derived from the $\xmax$ measured at the Yakutsk array is shown on the Fig.~\ref{fig6}. The $\MeanLnA$ calculations were performed with the use of the $\xmax$ predictions of QGSJet~II and SIBYLL models for proton and iron and with the relation~(\ref{eq3}) proposed by H\"{o}randel et al~\cite{Horandel2006}:

\begin{equation}
  \MeanLnA = \frac{\xmax - \xmax^{\text p}}{\xmax^{\text{Fe}} - \xmax^{\text{p}}} \cdot \ln{56}
  \label{eq3}
\end{equation}

The data from the Fig.~\ref{fig6} point towards the slight change of the $\MeanLnA$ value right after the kink in the spectrum. For example within the energy range $10^{16} - 10^{17}$~eV the $\MeanLnA$ value gets its maximum $\sim 3$ at $E_{0} = (5-8)\times 10^{16}$~eV (the composition becomes heavier) and above the energy $3 \times 10^{17}$~eV a decrease of the $\MeanLnA$ is pointed out (i.e. the composition becomes lighter). The data from other EAS arrays testify of the same tendency~\cite{Horandel2006}. If one plots the mass composition data versus the CR energy spectrum~(see Fig.~\ref{fig6}), then the coincidence between the peak of CR intensity and the maximum of the $\MeanLnA$ value could be clearly seen. Then it follows that the nature of the peak in the spectrum is related to a heavier component of cosmic rays.

\end{document}